\documentclass[aps,floatfix,showpacs,twocolumn]{revtex4}
\usepackage{amsmath}
\usepackage{amsfonts}
\usepackage{amssymb}
\usepackage{graphics,epsfig}
\usepackage{graphicx}
\begin{document}

\title{The volume of stationary black holes and the meaning of the surface gravity}
\author{William Ballik and Kayll Lake \cite{email}}
\affiliation{Department of Physics, Queen's University, Kingston,
Ontario, Canada, K7L 3N6 }
\date{\today}
\begin{abstract}
The invariant four-volume $\mathcal{V}$ of a complete black hole (the volume of the spacetime at and interior to the horizon) diverges. However, if one considers the black hole set up by the gravitational collapse of an object, and integrates only a finite time to the future of the collapse, the resultant volume is well defined and finite. In this paper we examine non-degenerate stationary black holes (and cosmological horizons) and find that  $\mathcal{V}_{s} \varpropto \ln(\lambda)$ where $s$ is any  shell that terminates on the horizon, $\lambda$ is the affine generator of the horizon and the constant of proportionality is the Parikh volume of $s$ divided by the surface gravity. This provides an alternative local and invariant definition of the surface gravity of a stationary black hole.
\end{abstract}
\maketitle
\section{Introduction}
\begin{figure}[ht]
\epsfig{file=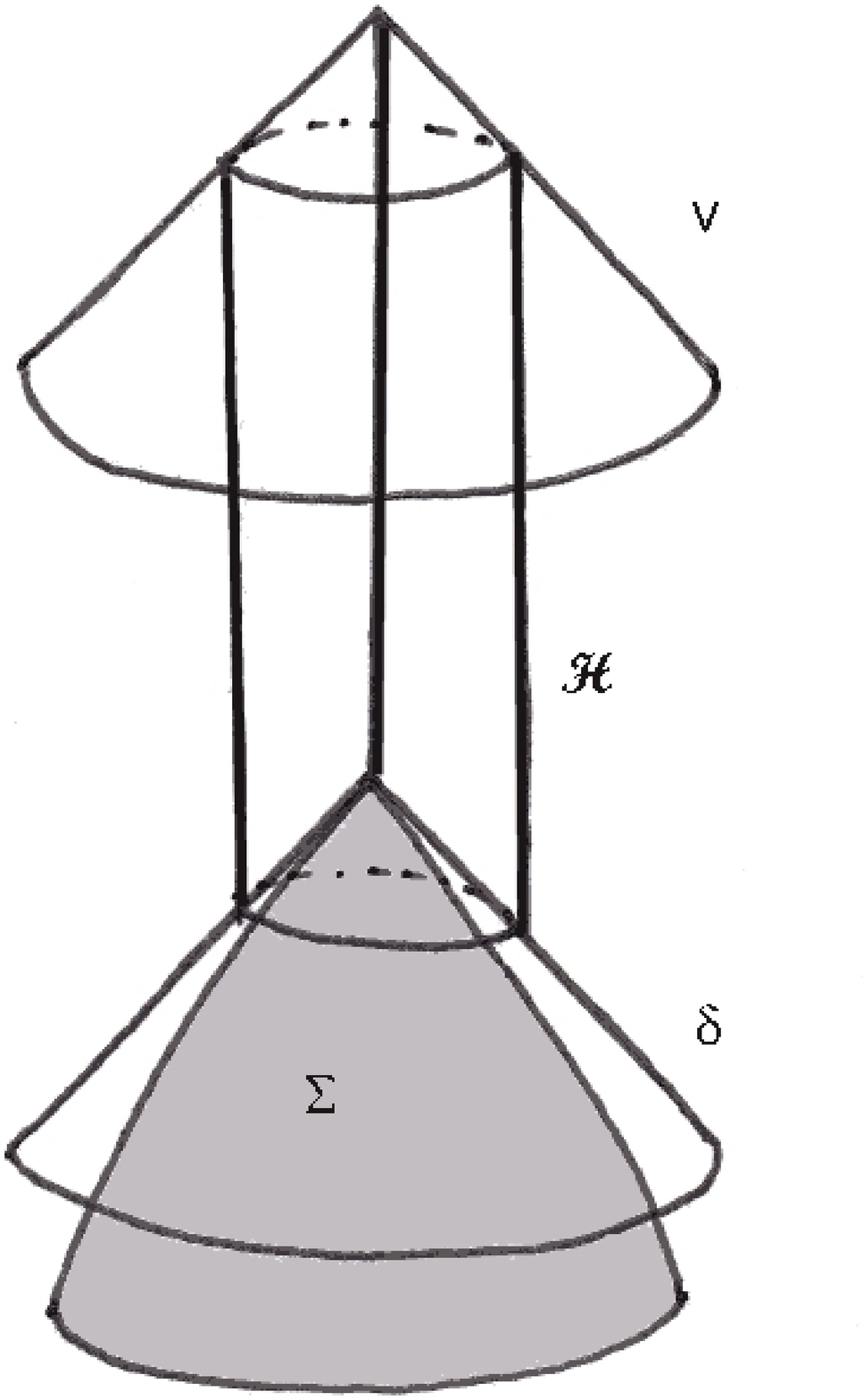,height=3in,width=2.7in,angle=0}
\caption{\label{figure}The collapse of a timelike boundary surface $\Sigma$ that terminates at the central singularity simultaneously with the null cone $\delta$ and produces a black hole with horizon $\mathcal{H}$. The null cone $v$ is any null cone to the future of $\delta$. The partial invariant four-volume $\mathcal{V}$ calculated here is bounded by $\delta$ and $v$, and is on and to the interior of $\mathcal{H}$.  We eventually show that the depth of the integration below $\mathcal{H}$ is irrelevant as regards the surface gravity $\kappa$. }
\end{figure}

The purpose of this paper is to discuss the importance of the rate of growth of the invariant four-volume $\mathcal{V}$ of stationary non-degenerate black holes and how this rate of growth is related to their surface gravity. Now whereas the three-volume of a black hole depends on the choice of slicing \cite{din}, and in general the associated three-volumes are finite, the full four-volume $\mathcal{V}$ is usually never discussed since it is formally infinite. (Consider, for example, the four-volume of the region on and below the future horizons in the Kruskal-Szekeres plane.) Even if we consider the black hole as created by the gravitational collapse of an object and so consider only part of the Kruskal-Szekeres plane on and below the future horizon and to the future of the boundary surface of the collapsing object, the four-volume still diverges as we integrate to the infinite future. However, we need not integrate to the infinite future and so in effect consider the evolution of $\mathcal{V}$. In this paper, to start, we examine explicitly in regular coordinates the four-volume bounded by the horizon, the central singularity and two distinct ingoing null cones in the Schwarzschild spacetime. The situation is shown schematically in Figure \ref{figure}. This introductory calculation points out a relation which we eventually show is a universal feature of stationary black holes: The invariant 4-volume $\mathcal{V}$ of a black hole grows as  $\mathcal{V} \varpropto \ln(\lambda)$ where $\lambda$ is the affine generator of the horizon and the constant of proportionality is the Parikh volume $^3\mathcal{V}$ \cite{parikh} (equivalent to the Euclidean 3-volume in spherical symmetry) divided by the surface gravity ($\kappa$). Defining $ \mathcal{V}^{*} \equiv d \mathcal{V} /d \ln(\lambda)$ then we have an alternative definition of the surface gravity: $\kappa=\;^3\mathcal{V}/\mathcal{V}^{*}$. However, the Reissner-Nordstr\"{o}m-de Sitter class of black holes, for example, contain unstable Cauchy horizons within their event horizons \cite{cauchy} and so one might argue that in such cases we would be integrating over unphysical regions of the spacetime. To overcome this we show that the four-volume of a shell ($s$) which terminates on the horizon is well defined and the shell can obviously be constructed so as to exclude the Cauchy horizons. In this case we find $\kappa=\;^3\mathcal{V}_{s}/\mathcal{V}^{*}_{s}$ and so $\kappa$ is in fact independent of the ``depth" of integration. We use such a construction to consider cosmological horizons and to argue that because of this independence on depth, $\kappa$ is a property of the horizon alone.

\bigskip

 The paper is organized as follows: First we give a general overview of the static spacetimes involved with our preliminary considerations. Then the invariant partial 4-volume of a Schwarzschild black hole is calculated explicitly in regular coordinates. To consider more general cases we first review the construction of regular coordinates for non-degenerate static black holes. We then calculate the invariant 4-volume $\mathcal{V}$ in this more general context and explain the shell construction necessary for the consideration of cosmological horizons. To move on to the stationary case we first review Parikh's construction and we then show that $\kappa=\;^3\mathcal{V}_{s}/\mathcal{V}^{*}_{s}$ holds in general. The explicit verification in the Kerr metric is relegated to an Appendix. We conclude with a discussion of this alternative definition of the surface gravity.
\section{Background}
The static spacetimes under consideration here can be written in the form
\begin{equation}\label{fform}
    ds^2=-fd t^2+\frac{d r^2}{f}+r^2d\Omega^2_{2}
\end{equation}
where $d\Omega^2_{2}$ is the metric of a unit two-sphere ($d\theta^2+\sin^2 \theta d \phi^2$) and $f=f(r)$, a polynomial with simple root(s) which locate the horizon(s). Within the context of classical general relativity the form (\ref{fform}) is of course well known as it includes the Reissner-Nordstr\"{o}m-de Sitter class of black holes. The form is also of interest for regular black holes (those without internal singularities), a subject that goes back many years \cite{conboy}. The curious appearance of (\ref{fform}) (that is, $g_{tt}g_{rr}=-1$) has been discussed recently by Jacobson \cite{jacobson} who showed, amongst other things, that the central feature of (\ref{fform}) is the vanishing radial null-null component of the Ricci tensor. Alternatively, consider the metric
\begin{equation}\label{jform}
    ds^2=-fd t^2+\frac{d r^2}{j}+r^2d\Omega^2_{2}
\end{equation}
where $j=j(r)$. Since $f$ is a polynomial with simple root(s) let us write
\begin{equation}\label{froot}
    f(r)=(r-a)h(r)
\end{equation}
where $h(a)\neq0$. Whereas for (\ref{fform}) all scalars constructed from the Riemann tensor are finite for $r>0$ and $h \in C^2$, for (\ref{jform})
these scalars diverge at $r=a$ unless $j(a)=0$. This, in some measure, helps to explain the prevalence of (\ref{fform}). However, the from (\ref{fform}) is defective at $r=a$ and so for our considerations of $\mathcal{V}$ we need regular coordinates.

\section{Schwarzschild}

We begin with a calculation of the complete invariant 4-volume of a Schwarzschild black hole in regular coordinates. As shown elsewhere \cite{lake}, the Kruskal \cite{kruskal} - Szekeres \cite{szekeres} form of the Schwarzschild metric can be given explicitly as
\begin{equation}\label{basemetric}
ds^2=(2M)^2 d\tilde{s}^2
\end{equation}
where
\begin{equation}\label{ks2}
d\tilde{s}^2=\frac{-4}{(1+\mathcal{L})e^{1+\mathcal{L}}}dudv+(1+\mathcal{L})^2d\Omega^2_{2}
\end{equation}
with $\mathcal{L}\equiv \mathcal{L}(-uv/e)$ where $\mathcal{L}$ is the Lambert W function \cite{lambert}.

Trajectories
with tangents \cite{tangents} $\mathcal{K}^{\alpha}=e^{\mathcal{L}}(1+\mathcal{L}) \delta^{\alpha}_{v}$ (constant $u=u_{0}, \theta$ and $\phi$) are radial null geodesics given by
\begin{equation}\label{u0}
    v(\lambda)=\lambda e^{-u_{0} \lambda /e}
\end{equation}
where $\lambda$ is an affine parameter (defined, of course, only up to a linear transformation) and we note the expansion
\begin{equation}\label{expK}
    \nabla_{\alpha}\mathcal{K}^{\alpha}=\frac{-2u_{0}}{e(1+\mathcal{L})}.
\end{equation}
Trajectories
with tangents $\mathcal{M}^{\alpha}=e^{\mathcal{L}}(1+\mathcal{L}) \delta^{\alpha}_{u}$ (constant $v=v_{0}, \theta$ and $\phi$) are radial null geodesics given by
\begin{equation}\label{v0}
    u(\lambda)=\lambda e^{-v_{0} \lambda /e}
\end{equation}
and we now note the expansion
\begin{equation}\label{expm}
    \nabla_{\alpha}\mathcal{M}^{\alpha}=\frac{-2v_{0}}{e(1+\mathcal{L})}.
\end{equation}
On the horizons $u=0$ and $v=0$ then $v$ and $u$ are affine parameters. The only singularity in (\ref{ks2}) occurs for $\mathcal{L}=-1$. That is, $uv=1$.

\bigskip

\begin{figure}[ht]
\epsfig{file=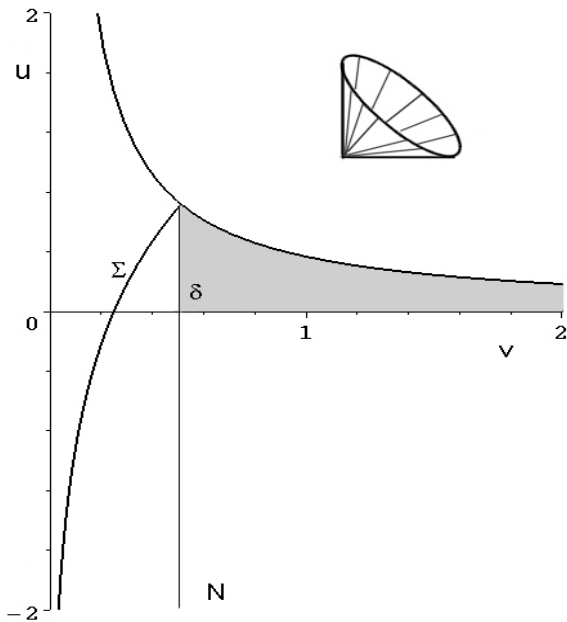,height=3.5in,width=3in,angle=0}
\caption{\label{figure1} $\mathcal{R}$ for the calculation of $\mathcal{V}$. The black hole is created by the collapse of some timelike boundary $\Sigma$. (The diagram to the left of $\Sigma$ is irrelevant.) The first ingoing null geodesic to hit the vacuum singularity is labeled $N$ and crosses the horizon at $v=\delta$.  }
\end{figure}

As is well known, the invariant 4-volume over some region $\mathcal{R}$ of spacetime is given by
\begin{equation}\label{volume}
    \mathcal{V}=\int \sqrt{|g|}\;^4dx
\end{equation}
where $g$ is the determinate of the metric and the integration is over $\mathcal{R}$. For the present calculation, $\mathcal{R}$ is defined in Figure \ref{figure1}.
From (\ref{volume}), (\ref{ks2}) and (\ref{basemetric}) we have
\begin{equation}\label{volumeK}
    \mathcal{V}=(2M)^4\int_{0}^{2 \pi}\int_{0}^{\pi}\int_{\delta}^{v}\int_{0}^{1/v}\frac{1+\mathcal{L}}{e^{1+\mathcal{L}}}\;du\; dv\; \sin(\theta)d \theta \;d \phi
\end{equation}
from which we find
\begin{equation}\label{answer}
    \mathcal{V}= \frac{8 \pi}{3}(2M)^4\ln(\frac{v}{\delta})
\end{equation}
so that with $u_{0}=0$ from (\ref{u0}) we arrive at the manifestly invariant statement
\begin{equation}\label{initial}
    \frac{d \mathcal{V}}{d \lambda}=\frac{8 \pi}{3}(2M)^4\frac{1}{\lambda},
\end{equation}
irrespective of initial conditions. This relation can be written in the equivalent form
\begin{equation}\label{finalvol}
    \frac{d \mathcal{V}}{d \lambda}=\frac{^3\mathcal{V}}{\kappa \lambda}
\end{equation}
where $\kappa \;(=\frac{1}{4M})$ is the surface gravity (see below) and $^3\mathcal{V} \;(=\frac{4 \pi}{3} (2M)^3)$ is the Euclidean 3-volume (also see below as the Euclidean form is due to the symmetry). Defining
\begin{equation}\label{vstar}
    \mathcal{V}^{*} \equiv \frac{d \mathcal{V} }{d \ln(\lambda)}
\end{equation}
we rewrite (\ref{finalvol}) as
\begin{equation}\label{kappaform}
    \kappa=\frac{^3\mathcal{V}}{\mathcal{V}^{*}}.
\end{equation}

\section{General $f$ - black holes}
We start with a known construction \cite{lake1} of regular coordinates for spherically symmetric static spacetimes with non-degenerate horizons.
\subsection{Construction of coordinates}
From (\ref{fform}) it follows that there exists coordinates $u$ and $v$ defined by
\begin{equation}\label{udef}
    2 \mathcal{C} \frac{du}{u}=\frac{dr}{f}-dt
\end{equation}
and
\begin{equation}\label{vdef}
     2 \mathcal{C} \frac{dv}{v}=\frac{dr}{f}+dt
\end{equation}
where $\mathcal{C}$ is a constant. The trajectories of constant $u$ (variable $v$) and of constant $v$ (variable $u$) label radial null geodesics. In particular, note that
\begin{equation}\label{utype}
    \mathcal{C}(\frac{du}{u}+\frac{dv}{v})=\frac{dr}{f}
\end{equation}
which, though obvious, is of particular use in what follows.
The intermediate form of the metric is
\begin{equation}\label{intermediate}
    ds^2=4\mathcal{C}^2f\frac{du dv}{u v}+r^2d\Omega^2_{2},
\end{equation}
where now $r=r(u,v)$. Since (\ref{intermediate}) remains defective at $r=a$, decompose $f$ as
\begin{equation}\label{decf}
   \frac{1}{f}=\frac{1}{h(a) (r-a)}+\frac{k(r)}{h(r)}
\end{equation}
where $k(a)/h(a) \neq 0$ and finite. It follows that
\begin{equation}\label{rtrans}
    uv=\pm(r-a)^{\frac{1}{\mathcal{C}h(a)}}\exp(\int \frac{k(r)}{\mathcal{C}h(r)}dr+\mathcal{E})
\end{equation}
where the sign depends on how we choose to orientate the $u-v$ axis and $\mathcal{E}$ is a constant.
We now have
\begin{equation}\label{doublenull}
    ds^2=\pm \frac{4 \mathcal{C}^2(r-a)h(r)}{(r-a)^{\frac{1}{\mathcal{C}h(a)}}}\exp(-\int \frac{k(r)}{\mathcal{C}h(r)}dr-\mathcal{E})dudv+r^2d\Omega^2_{2}
\end{equation}
where now, from (\ref{rtrans}),  $r=r(uv)$. From (\ref{doublenull}) we see that there is but one choice for $\mathcal{C}$ that gives a regular covering at $r=a$,
\begin{equation}\label{C}
    \mathcal{C}=\frac{1}{h(a)}=\frac{1}{2 \kappa}
\end{equation}
where we have defined the surface gravity \cite{surface}
\begin{equation}\label{surf}
    \kappa \equiv \frac{1}{2}\frac{df}{dr}\Big|_{a} > 0.
\end{equation}
Finally we note that
\begin{equation}\label{ttrans}
    \Big|\frac{v}{u}\Big|=\exp(2 \kappa t).
\end{equation}
In summary: under the transformations (\ref{rtrans}) and (\ref{ttrans}) the metric (\ref{fform}) takes the form
\begin{equation}\label{kruskal}
    ds^2=K(r)dudv+r^2d\Omega^2_{2}
\end{equation}
where again from (\ref{rtrans}), $r=r(uv)$ and
\begin{equation}\label{K}
    K(r)\equiv\pm \frac{a h(r)}{\kappa^2} \exp(-2 \kappa \int \frac{k(r)}{h(r)}dr).
\end{equation}
The integration constant has been absorbed into the factor $a$ and again the choice of sign determines the orientation of the $u-v$ axis. Under these transformations the Killing vector $\eta^{\alpha}=\delta^{\alpha}_{t}$ becomes $\bar{\eta}^{\alpha}=(u,-v,0,0).$ Note that the specified construction can always be done. However, about a distinct root, say $r= b \neq a$, a new chart must be constructed about $r=b$. With $f=1-2M/r$ we arrive back at (\ref{basemetric}) with (\ref{ks2}).
\subsection{Volume}
It is convenient to start with the intermediate form (\ref{intermediate}) so that with (\ref{C}) we have, prior to specifying boundary conditions,
\begin{equation}\label{intv}
    \mathcal{V}=4 \pi \int \left( \int \frac{f r^2}{2 \kappa u}du \right) \frac{dv}{\kappa v}.
\end{equation}
We now consider a shell ($s$) from $r=a$ to $r=r_{1}<a$ and make use of (\ref{utype}) and (\ref{C}) to write
\begin{equation}\label{intu}
    4 \pi \int \frac{f r^2}{2 \kappa u}du \rightarrow 4 \pi \int_{r_{1}}^ar^2 dr = \;^3\mathcal{V}_{s}
\end{equation}
and so from (\ref{intv}) and (\ref{intu})
\begin{equation}\label{ratev}
    \frac{d \mathcal{V}_{s}}{d v}=\frac{^3\mathcal{V}_{s}}{\kappa v}.
\end{equation}
As explained below, a shell-type construction is needed for the consideration of cosmological horizons.

To interpret $v$ we now use the regular form (\ref{kruskal}) with (\ref{K}). Trajectories
with tangents $\mathcal{K}^{\alpha}=\frac{-4}{K(r)} \delta^{\alpha}_{v}$ (constant $u=u_{0}, \theta$ and $\phi$) are radial null geodesics with expansion
\begin{equation}\label{expKg}
    \nabla_{\alpha}\mathcal{K}^{\alpha}=\frac{-8u_{0}}{K(r)r}\frac{dr}{d(uv)}.
\end{equation}
Trajectories
with tangents $\mathcal{M}^{\alpha}=\frac{-4}{K(r)} \delta^{\alpha}_{u}$ (constant $v=v_{0}, \theta$ and $\phi$) are radial null geodesics with expansion
\begin{equation}\label{expmg}
    \nabla_{\alpha}\mathcal{M}^{\alpha}=\frac{-8v_{0}}{K(r)r}\frac{dr}{d(uv)}.
\end{equation}
We set the horizons $r=a$ at $u=0$ and $v=0$ so that $v$ and $u$ are affine parameters. We can now rewrite (\ref{ratev}) in the form
\begin{equation}\label{ratevg}
    \frac{d \mathcal{V}_{s}}{d \lambda}=\frac{^3\mathcal{V}_{s}}{\kappa \lambda}.
\end{equation}
Note that section III above can be recovered form this section by setting $r_{1}=0$ and $f=1-2M/r$. Proceeding as above and defining
\begin{equation}\label{vstars}
    \mathcal{V}^{*}_{s} \equiv \frac{d \mathcal{V}_{s} }{d \ln(\lambda)}
\end{equation}
we rewrite (\ref{ratevg}) as
\begin{equation}\label{kappaform}
    \kappa=\frac{^3\mathcal{V}_{s}}{\mathcal{V}^{*}_{s}}.
\end{equation}
We see that $\kappa$ is in effect independent of the thickness of the shell $s$ and so we view it as a property of the horizon alone.
\section{General $f$ - cosmological horizons}
The foregoing argument applies directly to cosmological (and Cauchy) horizons with the insertion of an absolute value in the definition of the surface gravity in (\ref{surf}). We emphasize this point here by way of a direct calculation in de Sitter space in regular coordinates.

\bigskip

A complete covering of de Sitter space is given by
\begin{equation}\label{basedesitter}
        ds^2=\frac{3}{\Lambda}d\bar{s}^2
\end{equation}
where
\begin{equation}\label{desitter}
    d\bar{s}^2=\frac{-4 }{(1-uv)^2}du dv+ \left( \frac{1+uv}{1-uv} \right)^2 d \Omega_{2}^2
\end{equation}
and we consider the region $0<r \equiv \sqrt{\frac{3}{\Lambda}}(1+uv)/(1-uv)<\infty$.
Trajectories
with tangents $\mathcal{K}^{\alpha}=(1-u_{0}v)^2 \delta^{\alpha}_{v}$ (constant $u=u_{0}, \theta$ and $\phi$) are radial null geodesics given by
\begin{equation}\label{u0ds}
    v(\lambda)u_{0}=1-\frac{1}{u_{0}\lambda}
\end{equation}
where $\lambda$ is an affine parameter and $u_{0} \neq 0$. If $u_{0} = 0$ then the geodesic is affinely parameterized by $v$. We note the expansion
\begin{equation}\label{expKds}
    \nabla_{\alpha}\mathcal{K}^{\alpha}=4u_{0}\left(\frac{1-u_{0}v}{1+u_{0}v}\right).
\end{equation}
Trajectories
with tangents $\mathcal{M}^{\alpha}=(1-uv_{0})^2 \delta^{\alpha}_{u}$ (constant $v=v_{0}, \theta$ and $\phi$) are radial null geodesics given by
\begin{equation}\label{v0ds}
    u(\lambda)v_{0}=1-\frac{1}{v_{0} \lambda}
\end{equation}
for $v_{0} \neq 0$. If $v_{0} = 0$ then the geodesic is affinely parameterized by $u$. We now note the expansion
\begin{equation}\label{expmds}
    \nabla_{\alpha}\mathcal{M}^{\alpha}=4v_{0}\left(\frac{1-uv_{0}}{1+uv_{0}}\right).
\end{equation}
On the cosmological horizons $r=\sqrt{\frac{3}{\Lambda}}$, $u=0$ or $v=0$ and so $v$ and $u$ are affine parameters. There are no singularities in (\ref{desitter}). Note that $r=0$ for $uv=-1$ and $r=\infty$ for $uv=1$. To calculate a finite volume $\mathcal{V}$ for a cosmological horizon we integrate from  $r=\sqrt{\frac{3}{\Lambda}}$ out to (say) $\epsilon \sqrt{\frac{3}{\Lambda}}$ where $\epsilon > 1$. The situation considered is qualitatively similar to Figure \ref{figure1} but note that $r$ is now increasing along $N$. We now have
\begin{equation}\label{volumeKds}
    \mathcal{V}_{s}=4 \pi\left(\frac{3}{\Lambda}\right)^2\int_{\delta}^{v}\int_{0}^{\Delta/v}\frac{(1+uv)^2}{(1-uv)^4}\;du\; dv\
\end{equation}
where $0< \Delta \equiv (\epsilon-1)/(\epsilon+1)<1$. We find
\begin{equation}\label{answerds}
    \mathcal{V}_{s}= \frac{4 \pi}{3} \left(\frac{3}{\Lambda}\right)^2\left(\epsilon^3-1\right)\ln(\frac{v}{\delta})
\end{equation}
so that we arrive at the manifestly invariant statement
\begin{equation}\label{initialds}
    \frac{d\mathcal{V}_{s}}{d \lambda}=\frac{4 \pi}{3} \left(\frac{3}{\Lambda}\right)^2\left(\epsilon^3-1\right)\frac{1}{\lambda},
\end{equation}
irrespective of initial conditions. It is a simple matter to show that (\ref{initialds}) is equivalent to (\ref{ratevg}) and therefore (\ref{kappaform}).

\bigskip

All calculations up to this point can be considered merely motivational (but we think important) for the brief argument that now follows.
\section{The surface gravity $\kappa$}
Parikh \cite{parikh} has considered the volume
\begin{equation}\label{parikh}
   ^3\mathcal{V}^{*}=\int \sqrt{|g|}\;^3dx=\frac{d\mathcal{V}}{dT}
\end{equation}
for spacetimes with non-degenerate Killing vectors $\eta^{\alpha}$ with Killing parameter $T$ where $\eta^{\alpha}\nabla_{\alpha}T=1$. (Because the Parikh volume is a rate, we have introduced a superscript $*$ for consistency with (\ref{vstar}).)
It is important to note that the integrand in (\ref{parikh}) refers to the full spacetime and not a slice of it. For static spherically symmetric spacetimes it is easy to show that the Parikh volume at a horizon $r=a$ is simply the Euclidean 3-volume $4 \pi a^3/3$. However, away from (say) spherical symmetry, this will not be the case. Now let us write
\begin{equation}\label{fullkappa}
   \kappa \equiv \frac{^3\mathcal{V}^{*}_{s}}{\mathcal{V}^{*}_{s}}=\frac{1}{\lambda}\frac{d \lambda}{d T} \equiv \kappa.
\end{equation}
The right hand equivalence is a usual definition of the surface gravity $\kappa$ (see, for example, Wald \cite{surface}). The left hand equivalence is our interpretation of $\kappa$, as verified explicitly in the foregoing motivational calculations. Further explicit calculations seem essentially pointless in view of obvious equality in the center of (\ref{fullkappa}), and so we relegate the explicit verification in the Kerr metric to the Appendix.
\section{Discussion}
The usual physical meaning given to the surface gravity is, as explained for example by Poisson \cite{cauchy}, ``the force required of an observer at infinity to hold a particle (of unit mass) stationary at the horizon" (think of the Schwarzschild case). The interpretation given here, that the surface gravity is the ratio of the Parikh volume to the rate of change of the invariant four-volume for a shell of arbitrary (but non-vanishing) thickness bounded by the horizon, is a local interpretation that would appear to be new. An important question is, can we use this to gain further insights into black hole mechanics? The most obvious conclusion regards the third law of black hole mechanics, $\kappa \nrightarrow 0$. Since $^3\mathcal{V}_{s}>0$ even in the degenerate case (consider, for example, the static spherically symmetric case) we see that the third law demands that the rate of growth $\mathcal{V}^{*}_{s}$ must remain finite.
In order to violate the third law we need $\mathcal{V}^{*}_{s} \rightarrow \infty$ and since $d\mathcal{V}_{s}/d \lambda$ is finite, this requires $\lambda \rightarrow \infty$ in agreement with the formulation of Israel \cite{israel}. That is, in a sequence of quasi-static steps, the reduction of $\kappa$ to zero would take infinite advanced time.

\bigskip

\textit{Acknowledgments.} It is a pleasure to thank Maulik Parikh who pointed out \cite{parikh} the content of which changed our presentation from an earlier draft. This work was supported by a grant to KL from the Natural Sciences and Engineering Research Council of Canada. Portions of this work were made possible by use of \textit{GRTensorII} \cite{grt}.

\bigskip

\appendix*
\section{Kerr}\label{kerr}
In standard Kerr coordinates $(r,\theta,\phi,v)$ (e.g. Poisson \cite{cauchy} equation (5.65)) the null generator of the outer horizon ($r_{+}=m+\sqrt{m^2-a^2}$) can be given as
\begin{equation}\label{kerr}
   p^{\alpha}=(0,0,\frac{a}{\sqrt{m^2-a^2}}\exp(-\kappa v),\frac{1}{\kappa}\exp(-\kappa v))
\end{equation}
where
\begin{equation}\label{kappakerr}
    \kappa=\frac{\sqrt{m^2-a^2}}{2m(m+\sqrt{m^2-a^2})}.
\end{equation}
We have
\begin{equation}\label{kerrv}
    \mathcal{V}=\int_\frac{\ln \lambda_{1}}{\kappa}^\frac{\ln \lambda_{2}}{\kappa} \int_0^{2 \pi} \int_0^{\pi} \int_0^{r_{+}}sin(\theta)(r^2+a^2cos(\theta)^2)dr d \theta d \phi dv
\end{equation}
from which we obtain
\begin{equation}\label{finalkerr}
   \frac{d \mathcal{V}}{d \ln \lambda}=\frac{4}{3} \pi r_{+}(r_{+}^2+a^2)\frac{1}{\kappa}.
\end{equation}
We note that Parikh \cite{parikh} has already shown that
\begin{equation}\label{parikh1}
     ^3\mathcal{V}^{*}=\frac{4}{3} \pi r_{+}(r_{+}^2+a^2)
\end{equation}
and so with (\ref{finalkerr}) and (\ref{parikh1}) we arrive back at (\ref{fullkappa}) without $s$. To insert $s$ we simply integrate from $r_{0} < r_{+}$.


\begin{thebibliography}{}
\bibitem{tangents} All tangents, and associated expansions, are, of course, determined only up to a multaplicative constant. We include constants of interest.
\end{thebibliography}
\end{document}